\documentclass{article}
\usepackage{dcase2020,amsmath,amssymb,graphicx,url,times,booktabs, tabularx,glossaries,siunitx,color}
\usepackage{multirow}
\usepackage{pgfplots}
\pgfplotsset{compat=newest}
\usepackage{paralist}
\usepackage{balance}
\newacronym{FBCRNN}{FBCRNN}{forward-backward convolutional recurrent neural network}
\newacronym{CNN}{CNN}{convolutional neural network}
\newacronym{TCCNN}{TCCNN}{convolutional neural network}
\newacronym{DCASE}{DCASE}{detection and classification of acoustic scenes and events}
\newacronym{RNN}{RNN}{recurrent neural network}
\newacronym{SED}{SED}{sound event detection}
\newacronym{STFT}{STFT}{short-time Fourier transform}
\newacronym{SWA}{SWA}{stochastic weight averaging}
\newacronym{SOTA}{SOTA}{state-of-the-art}

\title{Forward-Backward Convolutional Recurrent Neural Networks and Tag-Conditioned Convolutional Neural Networks for \\Weakly Labeled Semi-supervised Sound Event Detection}

 \name{Janek Ebbers, Reinhold Haeb-Umbach\thanks{This work has been supported by Deutsche Forschungsgemeinschaft under contract no. HA 3455/15-1 within the Research Unit FOR 2457. Computational resources were provided by the Paderborn Center for Parallel Computing.}}
 \address{Paderborn University, Germany \\ \{ebbers, haeb\}@nt.upb.de}
\begin{document}

\ninept

\maketitle

\setlength{\abovedisplayskip}{4pt}
\setlength{\belowdisplayskip}{4pt}
\setlength{\abovedisplayshortskip}{0pt}
\setlength{\belowdisplayshortskip}{0pt}
\setlength{\textfloatsep}{5pt}
\renewcommand\arraystretch{1.0}
\setlength{\arraycolsep}{2pt}
\setlength{\abovecaptionskip}{0pt}
\setlength{\belowcaptionskip}{4pt}
\begin{sloppy}
\begin{abstract}
\vspace{-1mm}
In this paper we present our system for the \textit{\gls{DCASE} 2020 Challenge Task~4: Sound event detection and separation in domestic environments}.
We introduce two new models: the \gls{FBCRNN} and the tag-conditioned \gls{CNN}.
The \gls{FBCRNN} employs two \gls{RNN} classifiers sharing the same \gls{CNN} for preprocessing.
With one \gls{RNN} processing a recording in forward direction and the other in backward direction, the two networks are trained to jointly predict audio tags, i.e., weak labels, at each time step within a recording, given that at each time step they have jointly processed the whole recording.
The proposed training encourages the classifiers to tag events as soon as possible.
Therefore, after training, the networks can be applied to shorter audio segments of, e.g., \SI{200}{ms}, allowing \gls{SED}.
Further, we propose a tag-conditioned \gls{CNN} to complement \gls{SED}.
It is trained to predict strong labels while using (predicted) tags, i.e., weak labels, as additional input.
For training pseudo strong labels from a \gls{FBCRNN} ensemble are used.
The presented system scored the fourth and third place in the systems and teams rankings, respectively.
Subsequent improvements allow our system to even outperform the challenge baseline and winner systems in average by, respectively, \SI{18.0}{\%} and \SI{2.2}{\%} event-based $F_1$-score on the validation set.
Source code is publicly available at \url{https://github.com/fgnt/pb_sed}.
\end{abstract}
\vspace{-.5mm}
\begin{keywords}
\vspace{-2mm}
audio tagging, event detection, weak labels
\end{keywords}
\section{Introduction}
\label{sec:intro}
\vspace{-2mm}
\glsresetall
Environmental sound recognition is recently gaining increased interest from both academia and industry.
Plenty of applications potentially benefit from reliable sound recognition such as ambient assisted living, autonomous driving and environmental monitoring.
Depending on the application different acoustic information is required.
While acoustic scene classification aims at classifying the acoustic environment, audio tagging and \gls{SED} aim at recognizing specific sounds \cite{virtanen2018computational}.
The latter two differ in the provided level of detail with audio tagging only indicating the presence of a sound in a recording of, e.g., \SI{10}{s}, and \gls{SED} aiming at on- and offset detection within a certain collar of, e.g., \SI{200}{ms}.

With deep neural networks dominating \gls{SOTA} sound recognition, training requires labeled data, which, however, is expensive and time-consuming particularly with strong labels, i.e., when event on- and offsets have to be annotated.
Due to the additional effort for strong labeling, large scale databases like Google's AudioSet \cite{gemmeke2017audio} usually only provide weak labels which only indicate presence/absence of certain sounds within audio recordings.
Therefore, the first challenge in \gls{SED} is to learn to predict event on- and offsets despite the weak audio tagging labels provided during training.
Further, semi-supervised learning tries to only use few labeled data while exploiting unlabeled data to improve performance.

Driven by the annual \gls{DCASE} challenges \cite{Mesaros2017,Serizel2018,Turpault2019}, the \gls{SOTA} in weakly labeled semi-supervised \gls{SED} has progressed rapidly over the last years.
Several approaches have been proposed for weakly labeled \gls{SED} \cite{Adavanne2017,shah2018closer,mcfee2018adaptive,chou2018learning} most of which are based on multiple instance learning pooling functions \cite{wang2019comparison}.
Most recent \gls{SOTA} approaches, e.g., \cite{chou2018learning,yu2018multi,lin2020specialized,miyazaki2020weakly}, rely on neural attention.
To perform audio tagging a neural network learns to attend to the time range where the sound event is active.
Afterwards the network can be used to locate sound events in time although no strong labels have been used during training.
Semi-supervised \gls{SED} is dominated by teacher student approaches \cite{Lu2018,Lin2019}, where the teacher and student networks are jointly trained employing an additional loss for consistency between their predictions on unlabeled~data.

In this paper we present our system for the \textit{\gls{DCASE}~2020 Challenge Task~4: Sound event detection and separation in domestic environments} \cite{turpault2020a} tackling weakly labeled semi-supervised \gls{SED} with multi-label classification, i.e., multiple events can be active at a time.
It aims at \gls{SED} of ten different sound classes in real audio recordings from a domestic environment.
%The task features three training sets for \gls{SED}, namely a small weakly labeled data set of $1578$ real audio recordings, $2584$ synthetic soundscapes with strong labels and a larger set of $14412$ unlabeled real audio recordings.

We present a weakly labeled \gls{SED} approach based on two new models: the \gls{FBCRNN} and tag-conditioned \gls{CNN}.
The \gls{FBCRNN} employs a shared \gls{CNN} and two \glspl{RNN}, one processing the input audio signal in forward direction and the other in backward direction.
The \glspl{RNN} are encouraged to tag events as soon as possible by training them to jointly predict audio tags at each time step within a recording, given that at each time step the two \glspl{RNN} have jointly processed the whole recording.
After training, the networks can also be used for \gls{SED} by applying them to short audio segments of, e.g., \SI{200}{ms}.
As a complement, tag-conditioned \glspl{CNN} are trained to predict strong labels when getting tags as additional input.
Here, pseudo strong labels from a \gls{FBCRNN} ensemble are used for training.

It is shown that the proposed approach is highly competitive and outperforms current \gls{SOTA} approaches.
A rather naive pseudo labeling \cite{lee2013pseudo} of unlabeled data is shown to improve performance.
We hypothesize that a more sophisticated approach to semi-supervised learning, such as a mean-teacher approach \cite{Lu2018}, may even further improve performance in the future.

The rest of the paper is structured as follows.
Sec.~\ref{sec:feats} explains our feature extraction and data augmentation.
Then the \gls{FBCRNN} and tag-conditioned \gls{CNN} models are introduced in Sec.~\ref{sec:fbcrnn} and Sec.~\ref{sec:cnn}, respectively.
Finally, in Sec.~\ref{sec:exp} experiments are presented and conclusions are drawn in Sec.~\ref{sec:con}.

\section{Feature Extraction and Data Augmentation}
\label{sec:feats}
Our system's input $\mathbf{X}$ is a 128 dimensional log-mel spectrogram using a \gls{STFT} with a hop-size of \SI{20}{ms}, a frame length of \SI{60}{ms} and a sampling rate of \SI{16}{kHz}.
Waveforms are initially normalized to be within the range of -1 and 1: ${x(t) = s(t)/\max(|s(t)|)}$.
Each mel bin of the log-mel spectrogram is globally normalized to zero mean and unit variance.

During training various data augmentation techniques are used, namely random scaling, mixup, frequency warping, blurring, time masking, frequency masking and random noise.
Random scaling and mixup \cite{zhang2017mixup} are performed on the waveform similar as in \cite{Ebbers2019} by shifting and superposing signals as follows:
\begin{align*}
x'_i(t) &= \sum_{j=0}^{J_i-1} \lambda_j x_j(t-\tau_j)
\end{align*}
with the time shift $\tau_j$ being uniformly sampled such that $x'_i(t)$ is not longer than the maximum mixture length $T_\text{max}$, the mixture weights $\lambda_j$ being sampled from $\text{LogNormal}(0,1)$ and the distribution of the number of mixture components $J_i$ being
\begin{align*}
\Pr(J_i)=\begin{cases}
1-p_\text{mix}&J=1,\\
p_\text{mix}&J=2,\\
0&\mathrm{else}.\\
\end{cases}
\end{align*}
The mixup probability $p_\text{mix}$ and $T_\text{max}$ are hyper-parameters. 
Note the difference to original mixup \cite{zhang2017mixup} as we do not apply an interpolation to the signals but a superposition.
Therefore, we also do not interpolate the targets but combine them into a single multi-hot target vector.
Frequency warping and time- and frequency masking are performed exactly as in \cite{Ebbers2019}.
Blurring is performed with a gaussian blur kernel of size $5\times 5$ where the standard deviation of the kernel is randomly sampled from $\mathrm{Exp}(0.5)$.
Finally, random gaussian noise is added to the log-mel spectrogram with the noise power being randomly sampled from $\mathrm{Uniform}(0,0.2)$.

\section{Forward-Backward Convolutional Recurrent Neural Network}
\label{sec:fbcrnn}
To allow for \gls{SED} learned from weak labels, we aim to make a \gls{RNN} classifier (either processing an input signal in forward or backward direction) to immediately tag an event at the frame it appears.
If, however, a \gls{RNN} is only trained to predict tags after it has processed the whole signal, as in \cite{Ebbers2019}, there is no reason it should output tags~immediately.
More likely, it learns to gather information over long segments and not make hasty decisions.
To encourage the model to make immediate predictions, one might think to train a \gls{RNN} to predict event tags after each frame.
If an event, however, occurs only at the end of an audio-clip, the model would be trained to tag the event before it has seen it.
Here, we therefore suggest a joint training of a forward and backward \gls{RNN} such that at each frame at least one of the two \glspl{RNN} correctly tags an active event and both do not tag inactive events.
We hypothesize that forward and backward classifiers are able to jointly perform a tagging at each time frame as the forward classifier has seen all sound events between first and current frame while the backward classifier has seen all sound events between current and last frame.
This way we encourage the classifiers to tag sound events as soon as they appear.
Fig.~\ref{fig:fwd_bwd} illustrates the prediction behavior of the forward and backward classifiers.
With a shared \gls{CNN} as pre-processing we refer to the proposed model as \glsentryfull{FBCRNN}.
\begin{figure}[t]
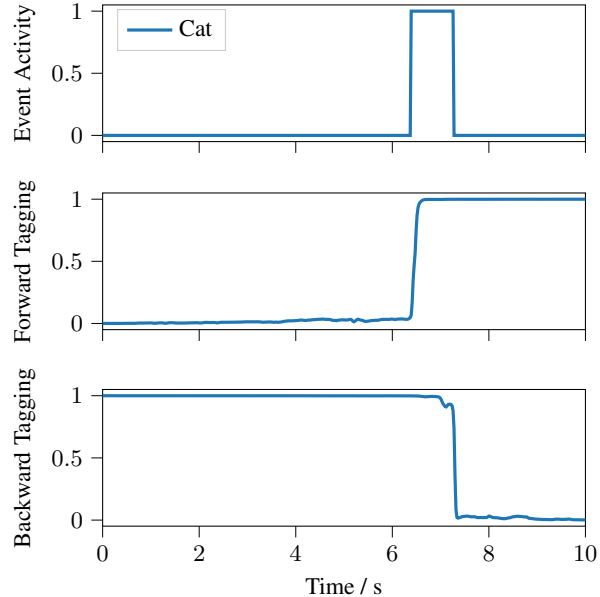

	\centering
    \newlength\figureheight
    \newlength\figurewidth
    \setlength\figureheight{3.4cm}
    \setlength\figurewidth{8cm}
	% This file was created by tikzplotlib v0.8.1.
\begin{tikzpicture}

\definecolor{color0}{rgb}{0.12156862745098,0.466666666666667,0.705882352941177}

\begin{axis}[
height=\figureheight,
legend cell align={left},
legend style={at={(0.03,0.97)}, anchor=north west, draw=white!80.0!black},
tick align=outside,
tick pos=left,
width=\figurewidth,
x grid style={white!69.01960784313725!black},
xmin=0, xmax=10,
xtick style={color=black},
xticklabels={{},{},{},{},{},{},\text{\quad\,\,\,}},
y grid style={white!69.01960784313725!black},
ylabel={Event Activity},
ymin=-0.05, ymax=1.05,
ytick style={color=black}
]
\addplot [very thick, color0]
table {%
0 0
0.0201612903225806 0
0.0403225806451613 0
0.0604838709677419 0
0.0806451612903226 0
0.100806451612903 0
0.120967741935484 0
0.141129032258065 0
0.161290322580645 0
0.181451612903226 0
0.201612903225806 0
0.221774193548387 0
0.241935483870968 0
0.262096774193548 0
0.282258064516129 0
0.30241935483871 0
0.32258064516129 0
0.342741935483871 0
0.362903225806452 0
0.383064516129032 0
0.403225806451613 0
0.423387096774194 0
0.443548387096774 0
0.463709677419355 0
0.483870967741935 0
0.504032258064516 0
0.524193548387097 0
0.544354838709677 0
0.564516129032258 0
0.584677419354839 0
0.604838709677419 0
0.625 0
0.645161290322581 0
0.665322580645161 0
0.685483870967742 0
0.705645161290323 0
0.725806451612903 0
0.745967741935484 0
0.766129032258065 0
0.786290322580645 0
0.806451612903226 0
0.826612903225806 0
0.846774193548387 0
0.866935483870968 0
0.887096774193548 0
0.907258064516129 0
0.92741935483871 0
0.94758064516129 0
0.967741935483871 0
0.987903225806452 0
1.00806451612903 0
1.02822580645161 0
1.04838709677419 0
1.06854838709677 0
1.08870967741935 0
1.10887096774194 0
1.12903225806452 0
1.1491935483871 0
1.16935483870968 0
1.18951612903226 0
1.20967741935484 0
1.22983870967742 0
1.25 0
1.27016129032258 0
1.29032258064516 0
1.31048387096774 0
1.33064516129032 0
1.3508064516129 0
1.37096774193548 0
1.39112903225806 0
1.41129032258065 0
1.43145161290323 0
1.45161290322581 0
1.47177419354839 0
1.49193548387097 0
1.51209677419355 0
1.53225806451613 0
1.55241935483871 0
1.57258064516129 0
1.59274193548387 0
1.61290322580645 0
1.63306451612903 0
1.65322580645161 0
1.67338709677419 0
1.69354838709677 0
1.71370967741935 0
1.73387096774194 0
1.75403225806452 0
1.7741935483871 0
1.79435483870968 0
1.81451612903226 0
1.83467741935484 0
1.85483870967742 0
1.875 0
1.89516129032258 0
1.91532258064516 0
1.93548387096774 0
1.95564516129032 0
1.9758064516129 0
1.99596774193548 0
2.01612903225806 0
2.03629032258065 0
2.05645161290323 0
2.07661290322581 0
2.09677419354839 0
2.11693548387097 0
2.13709677419355 0
2.15725806451613 0
2.17741935483871 0
2.19758064516129 0
2.21774193548387 0
2.23790322580645 0
2.25806451612903 0
2.27822580645161 0
2.29838709677419 0
2.31854838709677 0
2.33870967741935 0
2.35887096774194 0
2.37903225806452 0
2.3991935483871 0
2.41935483870968 0
2.43951612903226 0
2.45967741935484 0
2.47983870967742 0
2.5 0
2.52016129032258 0
2.54032258064516 0
2.56048387096774 0
2.58064516129032 0
2.6008064516129 0
2.62096774193548 0
2.64112903225806 0
2.66129032258065 0
2.68145161290323 0
2.70161290322581 0
2.72177419354839 0
2.74193548387097 0
2.76209677419355 0
2.78225806451613 0
2.80241935483871 0
2.82258064516129 0
2.84274193548387 0
2.86290322580645 0
2.88306451612903 0
2.90322580645161 0
2.92338709677419 0
2.94354838709677 0
2.96370967741935 0
2.98387096774194 0
3.00403225806452 0
3.0241935483871 0
3.04435483870968 0
3.06451612903226 0
3.08467741935484 0
3.10483870967742 0
3.125 0
3.14516129032258 0
3.16532258064516 0
3.18548387096774 0
3.20564516129032 0
3.2258064516129 0
3.24596774193548 0
3.26612903225806 0
3.28629032258065 0
3.30645161290323 0
3.32661290322581 0
3.34677419354839 0
3.36693548387097 0
3.38709677419355 0
3.40725806451613 0
3.42741935483871 0
3.44758064516129 0
3.46774193548387 0
3.48790322580645 0
3.50806451612903 0
3.52822580645161 0
3.54838709677419 0
3.56854838709677 0
3.58870967741935 0
3.60887096774194 0
3.62903225806452 0
3.6491935483871 0
3.66935483870968 0
3.68951612903226 0
3.70967741935484 0
3.72983870967742 0
3.75 0
3.77016129032258 0
3.79032258064516 0
3.81048387096774 0
3.83064516129032 0
3.8508064516129 0
3.87096774193548 0
3.89112903225806 0
3.91129032258065 0
3.93145161290323 0
3.95161290322581 0
3.97177419354839 0
3.99193548387097 0
4.01209677419355 0
4.03225806451613 0
4.05241935483871 0
4.07258064516129 0
4.09274193548387 0
4.11290322580645 0
4.13306451612903 0
4.15322580645161 0
4.17338709677419 0
4.19354838709677 0
4.21370967741935 0
4.23387096774194 0
4.25403225806452 0
4.2741935483871 0
4.29435483870968 0
4.31451612903226 0
4.33467741935484 0
4.35483870967742 0
4.375 0
4.39516129032258 0
4.41532258064516 0
4.43548387096774 0
4.45564516129032 0
4.4758064516129 0
4.49596774193548 0
4.51612903225806 0
4.53629032258065 0
4.55645161290323 0
4.57661290322581 0
4.59677419354839 0
4.61693548387097 0
4.63709677419355 0
4.65725806451613 0
4.67741935483871 0
4.69758064516129 0
4.71774193548387 0
4.73790322580645 0
4.75806451612903 0
4.77822580645161 0
4.79838709677419 0
4.81854838709677 0
4.83870967741935 0
4.85887096774194 0
4.87903225806452 0
4.8991935483871 0
4.91935483870968 0
4.93951612903226 0
4.95967741935484 0
4.97983870967742 0
5 0
5.02016129032258 0
5.04032258064516 0
5.06048387096774 0
5.08064516129032 0
5.1008064516129 0
5.12096774193548 0
5.14112903225806 0
5.16129032258065 0
5.18145161290323 0
5.20161290322581 0
5.22177419354839 0
5.24193548387097 0
5.26209677419355 0
5.28225806451613 0
5.30241935483871 0
5.32258064516129 0
5.34274193548387 0
5.36290322580645 0
5.38306451612903 0
5.40322580645161 0
5.42338709677419 0
5.44354838709677 0
5.46370967741935 0
5.48387096774194 0
5.50403225806452 0
5.5241935483871 0
5.54435483870968 0
5.56451612903226 0
5.58467741935484 0
5.60483870967742 0
5.625 0
5.64516129032258 0
5.66532258064516 0
5.68548387096774 0
5.70564516129032 0
5.7258064516129 0
5.74596774193548 0
5.76612903225806 0
5.78629032258065 0
5.80645161290323 0
5.82661290322581 0
5.84677419354839 0
5.86693548387097 0
5.88709677419355 0
5.90725806451613 0
5.92741935483871 0
5.94758064516129 0
5.96774193548387 0
5.98790322580645 0
6.00806451612903 0
6.02822580645161 0
6.04838709677419 0
6.06854838709677 0
6.08870967741935 0
6.10887096774194 0
6.12903225806452 0
6.1491935483871 0
6.16935483870968 0
6.18951612903226 0
6.20967741935484 0
6.22983870967742 0
6.25 0
6.27016129032258 0
6.29032258064516 0
6.31048387096774 0
6.33064516129032 0
6.3508064516129 0
6.37096774193548 0
6.39112903225806 1
6.41129032258065 1
6.43145161290323 1
6.45161290322581 1
6.47177419354839 1
6.49193548387097 1
6.51209677419355 1
6.53225806451613 1
6.55241935483871 1
6.57258064516129 1
6.59274193548387 1
6.61290322580645 1
6.63306451612903 1
6.65322580645161 1
6.67338709677419 1
6.69354838709677 1
6.71370967741935 1
6.73387096774194 1
6.75403225806452 1
6.7741935483871 1
6.79435483870968 1
6.81451612903226 1
6.83467741935484 1
6.85483870967742 1
6.875 1
6.89516129032258 1
6.91532258064516 1
6.93548387096774 1
6.95564516129032 1
6.9758064516129 1
6.99596774193548 1
7.01612903225806 1
7.03629032258065 1
7.05645161290323 1
7.07661290322581 1
7.09677419354839 1
7.11693548387097 1
7.13709677419355 1
7.15725806451613 1
7.17741935483871 1
7.19758064516129 1
7.21774193548387 1
7.23790322580645 1
7.25806451612903 1
7.27822580645161 0
7.29838709677419 0
7.31854838709677 0
7.33870967741935 0
7.35887096774194 0
7.37903225806452 0
7.3991935483871 0
7.41935483870968 0
7.43951612903226 0
7.45967741935484 0
7.47983870967742 0
7.5 0
7.52016129032258 0
7.54032258064516 0
7.56048387096774 0
7.58064516129032 0
7.6008064516129 0
7.62096774193548 0
7.64112903225806 0
7.66129032258065 0
7.68145161290323 0
7.70161290322581 0
7.72177419354839 0
7.74193548387097 0
7.76209677419355 0
7.78225806451613 0
7.80241935483871 0
7.82258064516129 0
7.84274193548387 0
7.86290322580645 0
7.88306451612903 0
7.90322580645161 0
7.92338709677419 0
7.94354838709677 0
7.96370967741935 0
7.98387096774194 0
8.00403225806452 0
8.0241935483871 0
8.04435483870968 0
8.06451612903226 0
8.08467741935484 0
8.10483870967742 0
8.125 0
8.14516129032258 0
8.16532258064516 0
8.18548387096774 0
8.20564516129032 0
8.2258064516129 0
8.24596774193548 0
8.26612903225806 0
8.28629032258064 0
8.30645161290323 0
8.32661290322581 0
8.34677419354839 0
8.36693548387097 0
8.38709677419355 0
8.40725806451613 0
8.42741935483871 0
8.44758064516129 0
8.46774193548387 0
8.48790322580645 0
8.50806451612903 0
8.52822580645161 0
8.54838709677419 0
8.56854838709677 0
8.58870967741935 0
8.60887096774194 0
8.62903225806452 0
8.6491935483871 0
8.66935483870968 0
8.68951612903226 0
8.70967741935484 0
8.72983870967742 0
8.75 0
8.77016129032258 0
8.79032258064516 0
8.81048387096774 0
8.83064516129032 0
8.8508064516129 0
8.87096774193548 0
8.89112903225806 0
8.91129032258064 0
8.93145161290323 0
8.95161290322581 0
8.97177419354839 0
8.99193548387097 0
9.01209677419355 0
9.03225806451613 0
9.05241935483871 0
9.07258064516129 0
9.09274193548387 0
9.11290322580645 0
9.13306451612903 0
9.15322580645161 0
9.17338709677419 0
9.19354838709677 0
9.21370967741935 0
9.23387096774194 0
9.25403225806452 0
9.2741935483871 0
9.29435483870968 0
9.31451612903226 0
9.33467741935484 0
9.35483870967742 0
9.375 0
9.39516129032258 0
9.41532258064516 0
9.43548387096774 0
9.45564516129032 0
9.4758064516129 0
9.49596774193548 0
9.51612903225806 0
9.53629032258064 0
9.55645161290323 0
9.57661290322581 0
9.59677419354839 0
9.61693548387097 0
9.63709677419355 0
9.65725806451613 0
9.67741935483871 0
9.69758064516129 0
9.71774193548387 0
9.73790322580645 0
9.75806451612903 0
9.77822580645161 0
9.79838709677419 0
9.81854838709677 0
9.83870967741935 0
9.85887096774194 0
9.87903225806452 0
9.8991935483871 0
9.91935483870968 0
9.93951612903226 0
9.95967741935484 0
9.97983870967742 0
10 0
};
\addlegendentry{Cat}
\end{axis}

\end{tikzpicture}
	\input{figures/forward.tex}
	\input{figures/backward.tex}
	\caption{Illustration of forward and backward tagging.}
	\label{fig:fwd_bwd}
%	\vspace{-3mm}
\end{figure}

%\subsection{Model}
%\label{sec:crnn_system}
%\vspace{-1mm}
An input $\mathbf{X}$ is forwarded through the \gls{CNN} $\mathbf{H} = f_\text{cnn}(\mathbf{X})$.
The \gls{CNN} architecture is shown in Tab. \ref{tab:cnn}.
We then perform recurrent forward tagging ${\mathbf{Y}^\text{fwd} = f_\text{rnn}^\text{fwd}(\mathbf{H})}$ and backward tagging ${\overleftarrow{\mathbf{Y}}^\text{bwd} = f_\text{rnn}^\text{bwd}(\overleftarrow{\mathbf{H}})}$, with $\overleftarrow{\mathbf{H}}$ denoting time flipped $\mathbf{H}$.
Note the difference to a bidirectional \gls{RNN} as here the forward and backward \glspl{RNN} do not exchange hidden representations.
The \gls{RNN} architecture is shown in Tab. \ref{tab:rnn}.
\begin{table}[t]
  \caption{CNN architecture as in \cite{Ebbers2019} but without temporal pooling. Each ConvXd uses a kernel size of three and a stride of one and includes BatchNorm \cite{ioffe2015batch} and ReLU.}
%\vspace{-1mm}
  \centering
  \begin{tabular}{c|c}
  Block & output shape\\
  \noalign{\hrule height 1.2pt}
  LogMel(128) & $1{\times}128{\times}N$\\
  GlobalNorm & $1{\times}128{\times}N$\\
  \noalign{\hrule height 1.2pt}
  2$\times$Conv2d(16) & $16{\times}128{\times}N$\\
  Pool2d($2{\times}1$) & $16{\times}64{\times} N$\\\hline
  2$\times$Conv2d(32) & $32{\times}64{\times} N$\\
  Pool2d($2{\times}1$) & $32{\times}32{\times} N$\\\hline
  2$\times$Conv2d(64) & $64{\times}32{\times} N$\\
  Pool2d($2{\times}1$) & $64{\times}16{\times} N$\\\hline
  2$\times$Conv2d(128) & $128{\times}16{\times} N$\\
  Pool2d($2{\times}1$) & $128{\times}8{\times} N$\\\hline
  Conv2d(256) & $256{\times}8{\times} N$\\
  Pool2d($2{\times}1$) & $256{\times}4{\times} N$\\
  Reshape & $1024{\times} N$\\\hline
  3$\times$Conv1d(256) & $256{\times} N$\\
  \end{tabular}
  \label{tab:cnn}
\end{table}
\begin{table}[t]
%\vspace{-2mm}
  \caption{Recurrent classifier architecture as in \cite{Ebbers2019}.}
%\vspace{-1mm}
  \centering
  \begin{tabular}{c|c}
  Block & output shape\\
  \noalign{\hrule height 1.2pt}
  2$\times$GRU(256) & $256{\times} N$\\
  \noalign{\hrule height 1.2pt}
  fc$_\text{ReLU}$(256) & $256{\times} N$\\
  fc$_\text{Sigmoid}$($K$) & $K{\times} N$\\
  \end{tabular}
  \label{tab:rnn}
\end{table}
%\vspace{-2mm}

%\subsection{Training}
%\vspace{-1mm}
During training tag predictions are computed at each time frame by taking the maximum of the forward and backward prediction ${\mathbf{Y}^\text{tag}=\max(\mathbf{Y}^\text{fwd}, \mathbf{Y}^\text{bwd})}$ with $\max(\cdot)$ denoting point wise maximum operation.

The training criterion is the binary cross entropy between tag predictions $\mathbf{y}^\text{tag}_n$ and the multi-hot target vector $\mathbf{z}^\text{tag}$:
\begin{align*}
L(\mathbf{y}^\text{tag}_n, \mathbf{z}^\text{tag}) = -\sum_{k=0}^{K-1}\Big(z^\text{tag}_k\log(y^\text{tag}_{n,k}) + (1-z^\text{tag}_k)\log(1- y^\text{tag}_{n,k})\Big)
\end{align*}
with $K=10$ denoting the number of target event classes.

%\subsection{Inference}
At inference time audio tag predictions are obtained as the average of the forward and backward predictions when they have processed the whole signal:
\begin{align*}
\hat{\mathbf{y}}^\text{tag}=f_\text{tag}(\mathbf{X})=(\mathbf{Y}^\text{fwd}[N-1] + \mathbf{Y}^\text{bwd}[0])/2
\end{align*}
with $N$ denoting the number of frames in $\mathbf{X}$.
Event-specific tagging thresholds $\alpha_k$ are used to get binary tag predictions
\begin{align*}
\hat{z}^\text{tag}_k = \big[\hat{y}^\text{tag}_k>\alpha_k\big] = \begin{cases}1, & \hat{y}^\text{tag}_k > \alpha_k,\\0,&\mathrm{else.}\end{cases}
\end{align*}
%For each event-class we choose the $\alpha_k$ giving the best audio tagging F$_1$-score on the validation set.

\gls{SED} is achieved by applying tagging to a small context around each frame: ${\hat{\mathbf{y}}^\text{sed}_n = f_\text{tag}(\mathbf{X}_{n-C_k:n+C_k})\cdot\hat{\mathbf{z}}^\text{tag}}$  with $n$ denoting the frame index and $C_k$ denoting an event-specific one-sided context length.
Again applying event-specific detection thresholds $\beta_k$ yields binary predictions $\hat{z}^\text{sed}_{n,k} = \big[\hat{y}^\text{sed}_{n,k}>\beta_k\big]$.
Finally median filtering with an event-specific filter size $M_k$ is applied to get the final \gls{SED}.

\section{Tag-Conditioned Convolutional Neural Network}
\label{sec:cnn}
Second, we propose a tag-conditioned \gls{CNN} to perform \gls{SED} by directly predicting strong labels.
This model can be understood as a two-stage approach, where tags are predicted in a first stage, here by the \gls{FBCRNN}, and given the tag predictions a subsequent model, here a \gls{CNN}, predicts strong labels.
We hypothesize that the \gls{CNN} can do better event detection if it is aware of the active events within a recording, i.e. when it has to predict $\Pr(\text{event active in frame}|\text{event active in recording})$ rather than $\Pr(\text{event active in frame})$.

The \gls{CNN} architecture $\hat{\mathbf{y}}^\text{sed}_n=f_\text{sed}(\mathbf{X}_{n-R:n+R}, \mathbf{z}^\text{tag})$ is similar as in Tab.~\ref{tab:cnn} with $R=13$ being the one-sided receptive field of the \gls{CNN}.
Hence, the overall receptive field is $2R+1=27$ frames corresponding to \SI{580}{ms}.
In contrast to Tab.~\ref{tab:cnn} the last Conv1d layer is outputting $K=10$ scores here, one for each event class.
Further, in addition to the (augmented) log-mel input spectrogram, this second-stage \gls{CNN} is conditioned on audio tags by concatenating a multi-hot tag encoding $\mathbf{z}^\text{tag}$ to each time-frequency bin along the channel dimension of the log-mel spectrogram as well as to the hidden representation between the reshape operation and the first Conv1d layer.
%This allows the \gls{CNN} to search for frequency-local as well as overall patterns to recognize the activity of a predicted tag in a frame.

At training time (pseudo) weak labels $\mathbf{z}^\text{tag}$ are used as conditioning.
The training criterion is the frame-wise binary cross-entropy between predictions $\hat{\mathbf{y}}^\text{sed}_n$ and (pseudo) strong labels $\mathbf{z}^\text{sed}_n$:
\begin{align*}
L(\hat{\mathbf{y}}^\text{sed}_n, \mathbf{z}^\text{sed}_n) = -{\sum_{k=0}^{K-1}\Big(z^\text{sed}_{n,k}\log(\hat{y}^\text{sed}_{n,k}) + (1{-}z^\text{sed}_{n,k})\log(1{-}\hat{y}^\text{sed}_{n,k})\Big)}
\end{align*}
which is averaged over all frames in a mini-batch.
If no strong labels are available, pseudo strong labels from some other weakly trained \gls{SED} model, such as the \gls{FBCRNN}, can be used for training.

At inference time the CNN is conditioned on tag predictions~$\hat{\mathbf{z}}^\text{tag}$ from the \gls{FBCRNN}.
We apply event-wise decision thresholds~$\beta_k$, as before, to get binary frame predictions ${\hat{z}^\text{sed}_{n,k} = \big[\hat{y}^\text{sed}_{n,k}>\beta_k\big]}$ and subsequent median filtering with an event-specific filter size~$M_k$.

%However, to prevent the \gls{CNN} from simply forwarding the tags we add some noise $\boldsymbol{\epsilon}$ to the tags $\mathbf{z}$  during training yielding $\mathbf{z}'=\mathbf{z} + \boldsymbol{\epsilon}$.
%The noise is chosen such that $\Pr(z_k=1|z'_k=1)\approx 1/2$.

\section{Experiments}
\label{sec:exp}
Experiments are performed using the DESED database used in the fourth task of the \textit{\gls{DCASE}}~2019 and \textit{\gls{DCASE}}~2020 Challenges \cite{Turpault2019,turpault2020a}.
The database features three training sets for \gls{SED}, namely a small weakly labeled data set of $1578$ real audio recordings, $2584$ synthetic soundscapes with strong labels and a larger set of $14412$ unlabeled real audio recordings.
To adjust the percentage of the different data sets in training, in each epoch recordings from the weakly labeled and synthetic data sets are presented 10 and 2 times, respectively, resulting in a data distribution of \SI{75.3}{\%} weakly labeled and \SI{24.7}{\%} synthetic data.
In experiments using pseudo labeled recordings from the unlabeled data set, these recordings are presented once in each epoch resulting in a data distribution of \SI{44.6}{\%} weakly labeled, \SI{14.6}{\%} synthetic and \SI{40.8}{\%} unlabeled data.
For the synthetic data, we further perform on-the-fly reverberation of individual sound events.

All trainings are performed for $40000$ update steps with checkpointing and validation every $1000$-th update step.
Mini-batches of size $B$ are randomly sampled from the training data such that no signal in the mini batch is padded by more than 5\% and each mini-batch includes at least $\lfloor B/3 \rfloor$ examples from the weakly labeled data set.
Adam \cite{kingma2014adam} is used for optimization with gradient clipping at a threshold of $20$ and a learning rate ramp up to $5{\cdot} 10^{-4}$ over the first $1000$ update steps and a learning rate reduction to $1{\cdot}10^{-4}$ for update steps ${>}15000$.
While for \glspl{FBCRNN} the checkpoint with the best macro-averaged audio tagging F$_1\text{-}$score on the validation set is adopted as the final model, for \gls{CNN} models the checkpoint with the best macro-averaged frame-based F$_1\text{-}$score is chosen.

Reported audio tagging performance is the macro-averaged F$_1\text{-}$score and reported \gls{SED} performance is the macro-averaged event-based F$_1$-score \cite{mesaros2016metrics} using an onset collar of \SI{200}{ms} and an offset collar of \SI{200}{ms} or \SI{20}{\%} of event duration if the duration is~${>}\SI{1}{s}$.
If not stated otherwise, the hyper-parameters $\alpha_k$, $\beta_k$, $C_k\in\{5,10,15,20\}$ and $M_k\in\{11,21,31,41,51\}$ are tuned to give best performance on the validation set.
Since the labels of the \gls{DCASE} 2020 Task~4 evaluation set (eval-2020) are not public yet, performance is primarily evaluated on the publicly available \textit{\gls{DCASE}~2019 Task~4 youtube evaluation set} (yt-eval-2019) \cite{turpault_nicolas_2019_3588172,Turpault2019}.
If not stated otherwise, each experiment is repeated five times from which we report the mean and standard deviation.

Ensembles combine four independently trained models by averaging their model outputs $\hat{\mathbf{y}}$ if not stated otherwise.

First, we evaluate the single-model tagging and detection performance of the proposed \gls{FBCRNN}.
In particular the effectiveness of the proposed forward-backward approach for weakly labeled learning is investigated as well as the usefulness of pseudo labeling the unlabeled data for semi-supervised learning.
Tab.~\ref{tab:results-fbcrnn} compares the following models:
\begin{itemize}
\item CRNN$^\text{last-only}_\text{no-pseudo}$: CRNN w/ only forward tagging where loss is only computed at the last frame of an audio recording (see~\cite{Ebbers2019}) and trained w/o unlabeled data set,
\item CRNN$_\text{no-pseudo}$: CRNN w/ only forward tagging where loss is computed at each frame of an audio recording and trained w/o unlabeled data set,
\item FBCRNN$_\text{no-pseudo}$: \gls{FBCRNN} trained w/o unlabeled data set,
\item FBCRNN$_\text{submitted}$: \gls{FBCRNN} used in our submission trained w/ a heuristical on-the-fly pseudo labeling  \cite{Ebbers2020} (means and standard deviations computed over only four models here),
\item FBCRNN: \gls{FBCRNN} trained w/ unlabeled data set after weakly pseudo labeled by an FBCRNN$_\text{no-pseudo}$ ensemble.
\end{itemize}
The models are trained using a mini-batch size of $B=16$ and mixup with $p_\text{mix}=2/3$ and $T_\text{max}=\SI{15}{s}$.

\begin{table}[t]
	\caption{Single-model tagging and detection performance of CRNNs in terms of macro-averaged (event-based) F$_1$-scores[\%].}
	\vspace{-1mm}
	\label{tab:results-fbcrnn}
	\centering
	\setlength{\tabcolsep}{2pt}
	\begin{tabular}{l|ccccc}
	\multirow{2}{*}{\qquad Model}& \multicolumn{2}{c}{validation} & \multicolumn{2}{c}{yt-eval-2019}\\
		& Tagging & Detection & Tagging & Detection\\
  \noalign{\hrule height 1.2pt}
		CRNN$^\text{last-only}_\text{no-pseudo}$ & 81.8{\scriptsize${\pm}$0.2} & 25.5{\scriptsize${\pm}$0.6} & 80.9{\scriptsize${\pm}$0.7} & 18.1{\scriptsize${\pm}$0.5} \\  % /net/vol/ebbers/exp/dcase20-eval/2020-07-06-08-27-21
		CRNN$_\text{no-pseudo}$ & 79.6{\scriptsize${\pm}$0.5} & 34.0{\scriptsize${\pm}$0.7} & 78.7{\scriptsize${\pm}$0.4} & 30.9{\scriptsize${\pm}$1.4} \\
		FBCRNN$_\text{no-pseudo}$ & 82.6{\scriptsize${\pm}$0.1} & 40.7{\scriptsize${\pm}$1.3} & 81.8{\scriptsize${\pm}$0.5} & 40.3{\scriptsize${\pm}$1.9}  \\ 
		FBCRNN$_\text{submitted}$ & 82.3{\scriptsize${\pm}$0.4} & 42.0{\scriptsize${\pm}$0.2} & 81.8{\scriptsize${\pm}$0.5} & 41.3{\scriptsize${\pm}$1.1}  \\
		FBCRNN & 84.8{\scriptsize${\pm}$0.2} & 46.4{\scriptsize${\pm}$0.5} & 84.1{\scriptsize${\pm}$1.2} & 47.4{\scriptsize${\pm}$1.1}\\  %/net/vol/ebbers/exp/dcase20-eval/2020-07-04-17-27-00
	\end{tabular}
\vspace{-1mm}
\end{table}
%\begin{table}[h]
%	\caption{Event-based F$_1$-scores[\%] of CRNN ensembles.}
%%	\vspace{-1mm}
%	\label{tab:results-fbcrnn}
%	\centering
%	\setlength{\tabcolsep}{2pt}
%	\begin{tabular}{l|ccccc}
%	\multirow{2}{*}{\,\,\,\, Ensemble}& \multicolumn{2}{c}{validation} & \multicolumn{2}{c}{yt-eval-2019}\\
%		& Tagging & Detection & Tagging & Detection\\
%  \noalign{\hrule height 1.2pt}
%		CRNN$^\text{last-only}_\text{no-pseudo}$ & 82.9 & 27.2 & 82.0 & 15.9 \\  % /net/vol/ebbers/exp/dcase20-eval/2020-07-06-08-27-21
%		CRNN$_\text{no-pseudo}$ &&&&\\
%		$\text{FBCRNN}_\text{no-pseudo}$ & 83.9 & 43.0 & 83.8 & 46.1 \\  % /net/vol/ebbers/exp/dcase20-eval/2020-07-05-18-09-45
%		$\text{FBCRNN}_\text{submitted}$ & 83.7 & 44.3 & 82.9 & 46.8 \\  % /net/vol/ebbers/exp/dcase20-eval/2020-07-05-18-51-12
%		FBCRNN & 86.0 & 47.5 & 84.4 & 50.0\\  %/net/vol/ebbers/exp/dcase20-eval/2020-07-04-17-27-00
%	\end{tabular}
%%\vspace{-1mm}
%\end{table}

It can be observed that the CRNN$^\text{last-only}_\text{no-pseudo}$ gives good audio tagging performance but fails to perform \gls{SED}.
This confirms our hypothesis from Sec.~\ref{sec:fbcrnn} that without a frame-wise loss the model does not learn to tag sounds in short contexts.
Using a frame-wise loss with the forward-only CRNN (CRNN$_\text{no-pseudo}$) improves \gls{SED} to some extent.
However, training the model to tag events at frames, where it may not have seen the events yet, limits performance as can be seen by the deterioration of tagging.
Using the proposed forward backward tagging approach (FBCRNN$_\text{no-pseudo}$) significantly improves \gls{SED} performance.
Interestingly it also improves audio tagging performance over the CRNN$^\text{last-only}_\text{no-pseudo}$ model.
Finally, the on-the-fly pseudo labeling ($\text{FBCRNN}_\text{submitted}$), that was used in our submitted system, only slightly improves detection performance over the model trained without unlabeled data ($\text{FBCRNN}_\text{no-pseudo}$).
However, pseudo labeling using a $\text{FBCRNN}_\text{no-pseudo}$ ensemble and training a new FBCRNN by also leveraging pseudo labeled data, improves tagging and detection performance significantly.

Next the effectiveness of the tag-conditioned \gls{CNN} is evaluated.
For that we compare \glspl{CNN} with and without tag conditioning.
For training strong pseudo labels for the weakly and unlabeled data sets are obtained by an FBCRNN ensemble using hyper parameters giving best frame-based $F_1$-scores on the validation set.
The FBCRNN ensemble also provides the tags for conditioning.
The models are trained using a mini-batch size of $B=24$ and mixup with $p_\text{mix}=1/2$ and $T_\text{max}=\SI{12}{s}$.
Tab. \ref{tab:results-cnn} compares \gls{SED} performance of the two models.
First, it can be noted that both \glspl{CNN} improve detection performance over the FBCRNN.
This suggests that pseudo strong label training may improve performance in general.
Comparing the CNN models to each other, it can be seen that the tag conditioning improves average performance from 50.1\% to 53.4\% event-based $F_1$-score on the evaluation set.

Finally, ensemble performance is compared to challenge baseline and winner systems.
Tab. \ref{tab:results-final} lists detection performance for the following systems:
\begin{compactitem}
\item Baseline: baseline system 2020 \cite{turpault2020a},
\item Winner2019: winner system 2019 \cite{Lin2019},
\item Winner2020: winner system 2020 \cite{Miyazaki2020},
\item Hybrid$_\text{submitted}$: Our submitted system consisting of the four FBCRNN$_\text{submitted}$ and four tag conditioned CNNs trained on pseudo strong labels from FBCRNN$_\text{submitted}$ and where for hyper-parameter tuning only ${C_k\in\{5,10\}}$ and ${M_k\in\{21,41\}}$ have been considered \cite{Ebbers2020},
\item Hybrid$^*_\text{submitted}$: Models from Hybrid$_\text{submitted}$ with the more extensive hyper-parameter tuning,
\item FBCRNN: FBCRNN ensemble,
\item CNN: CNN ensemble w/ tag conditioning,
\item Hybrid: Combination of FBCRNN and CNN ensembles,
\end{compactitem}

\begin{table}[t]
	\caption{Single-model detection performance of CNN models in terms of macro-averaged event-based F$_1$-scores[\%].}
	\vspace{-1mm}
	\label{tab:results-cnn}
	\centering
	\setlength{\tabcolsep}{2pt}
	\begin{tabular}{l|ccc}
	\quad Model & validation & yt-eval-2019 \\
  \noalign{\hrule height 1.2pt}
		CNN$_\text{no-cond.}$ & 49.4{\scriptsize${\pm}$0.4} & 50.1{\scriptsize${\pm}$0.2} \\
		CNN & 51.5{\scriptsize${\pm}$0.7} & 53.4{\scriptsize${\pm}$0.7} \\
	\end{tabular}
\vspace{-1mm}
\end{table}
\begin{table}[t]
	\caption{Ensemble detection performance in terms of macro-averaged event-based F$_1$-scores[\%].}
	\vspace{-1mm}
	\label{tab:results-final}
	\centering
	\setlength{\tabcolsep}{2pt}
	\begin{tabular}{l|ccc}
	\quad Ensemble & validation & yt-eval-2019 & eval-2020\\
  \noalign{\hrule height 1.2pt}
		Baseline & 34.8\phantom{\scriptsize${\pm}$0.0} & -\phantom{\scriptsize${\pm}$0.0} & 34.9\phantom{\scriptsize${\pm}$0.0} \\
		Winner2019 & 45.3\phantom{\scriptsize${\pm}$0.0} & 47.7\phantom{\scriptsize${\pm}$0.0} & -\phantom{\scriptsize${\pm}$0.0} \\
		Winner2020 & 50.6\phantom{\scriptsize${\pm}$0.0} & -\phantom{\scriptsize${\pm}$0.0} & 51.1\phantom{\scriptsize${\pm}$0.0} \\
  \noalign{\hrule height 0.2pt}
		Hybrid$_\text{submitted}$ & 48.3\phantom{\scriptsize${\pm}$0.0} & 50.8\phantom{\scriptsize${\pm}$0.0} & 47.2\phantom{\scriptsize${\pm}$0.0} \\  % /net/vol/ebbers/exp/dcase20-eval
		Hybrid$^*_\text{submitted}$ & 49.2\phantom{\scriptsize${\pm}$0.0} & 51.1\phantom{\scriptsize${\pm}$0.0} & -\phantom{\scriptsize${\pm}$0.0} \\  % /net/vol/ebbers/exp/dcase20-eval
		FBCRNN & 48.3{\scriptsize${\pm}$0.4} & 49.9{\scriptsize${\pm}$1.1} & -\phantom{\scriptsize${\pm}$0.0} \\  %/net/vol/ebbers/exp/dcase20-eval/2020-07-04-17-27-00
		CNN & 52.1{\scriptsize${\pm}$0.5} & 54.1{\scriptsize${\pm}$0.7} & -\phantom{\scriptsize${\pm}$0.0} \\  % /net/vol/ebbers/exp/dcase20-eval/2020-07-05-18-48-24
		Hybrid & 52.8{\scriptsize${\pm}$0.6} & 54.6{\scriptsize${\pm}$0.5} & -\phantom{\scriptsize${\pm}$0.0} \\  % /net/vol/ebbers/exp/dcase20-eval/2020-07-05-16-00-50
	\end{tabular}
\end{table}

Ensembling gives on yt-eval-2019 an average gain of 2.4\% and 0.7\% event-based $F_1$-score over single-model \glspl{FBCRNN} and \glspl{CNN} (Tab.~\ref{tab:results-fbcrnn} and Tab.~\ref{tab:results-cnn}), respectively.
It can be noted that the \gls{CNN} ensemble clearly outperforms the baseline, winner and our submitted system.
Combining the four \glspl{FBCRNN} and four \glspl{CNN} from individual ensembles to a larger Hybrid ensemble of eight sub-models yields another slight improvement of 0.5\% over the \gls{CNN}-only ensemble.
However, do note that the \gls{CNN}-only detection is computationally much more efficient, as the \gls{FBCRNN} detection relies on processing a context of length $1+2C$ at each frame for all $C\in\{C_k\}_{k=1}^K$.
Comparing Hybrid$_\text{submitted}$ and Hybrid$^*_\text{submitted}$, which only differ in the hyper-parameters $\alpha_k$,$\beta_k$,$C_k$ and $M_k$, shows that the more extensive hyper-parameter tuning improves performance only slightly.
Therefore, the improvement of the CNN and Hybrid ensembles over our submitted system comes mainly due to the improved pseudo-labeling resulting in better FBCRNN performance, as already seen in Tab.~\ref{tab:results-fbcrnn}, and consequently in better pseudo strong labels for \gls{CNN} training and finally in better overall performance.
Tab.~\ref{tab:results} shows event wise performance of our proposed ensembles on yt-eval-2019 with bold values highlighting the best ensemble.

\vspace{-1mm}
\begin{table}[h]
  \caption{Event-wise detection performance on yt-eval-2019 in terms of event-based F$_1$-scores[\%].}
	\vspace{-1mm}
  \centering
  \begin{tabular}{c|ccc}
  Event & FBCRNN & CNN & Hybrid\\
  \noalign{\hrule height 1.2pt}
  Alarm bell ringing & 30.0{\scriptsize${\pm}$1.3} & \textbf{45.4{\scriptsize${\pm}$3.6}} & 44.8{\scriptsize${\pm}$1.8} \\
  Blender & 50.4{\scriptsize${\pm}$4.3} & 54.9{\scriptsize${\pm}$2.8} & \textbf{56.5{\scriptsize${\pm}$4.1}} \\
  Cat & 67.3{\scriptsize${\pm}$1.9} & \textbf{69.7{\scriptsize${\pm}$3.5}} & 68.7{\scriptsize${\pm}$2.7}\\
  Dishes & 31.2{\scriptsize${\pm}$1.8} & 29.5{\scriptsize${\pm}$2.9} & \textbf{32.1{\scriptsize${\pm}$3.9}}\\
  Dog & \textbf{48.3{\scriptsize${\pm}$1.6}} & 44.1{\scriptsize${\pm}$2.5} & 48.2{\scriptsize${\pm}$3.3} \\
  E. shaver/toothbrush & 44.6{\scriptsize${\pm}$1.8} & \textbf{52.5{\scriptsize${\pm}$3.0}} & 49.2{\scriptsize${\pm}$6.4} \\
  Frying & 58.5{\scriptsize${\pm}$1.0} & 61.4{\scriptsize${\pm}$1.1} & \textbf{61.7{\scriptsize${\pm}$1.2}} \\
  Running water & 39.7{\scriptsize${\pm}$4.1} & 45.7{\scriptsize${\pm}$2.7} & \textbf{46.6{\scriptsize${\pm}$2.1}}\\
  Speech & 60.8{\scriptsize${\pm}$1.4} & 64.0{\scriptsize${\pm}$0.8} & \textbf{64.1{\scriptsize${\pm}$1.1}} \\
  Vacuum cleaner & 68.4{\scriptsize${\pm}$1.1} & 73.5{\scriptsize${\pm}$1.1} & \textbf{73.8{\scriptsize${\pm}$1.8}} \\
  \end{tabular}
  \label{tab:results}
\vspace{-2mm}
\end{table}

\vspace{-2mm}
\section{Conclusions}
\label{sec:con}
\vspace{-1mm}
In this paper we introduced the \gls{FBCRNN}, a weakly labeled \gls{SED} model trained to perform forward and backward tagging.
The proposed training allows the model to also be applied to small segments of only a few hundred milliseconds enabling \gls{SED}.
On top we proposed a \gls{CNN}, which is conditioned on audio tags, as a complementary \gls{SED} model.
Our system scored the fourth and third place in the systems and teams ranking, respectively, of the \gls{DCASE}~2020 Challenge Task~4.
Subsequent improvements allow our system to even outperform the baseline and winner systems in average by, respectively, \SI{18.0}{\%} and \SI{2.2}{\%} event-based $F_1$-score on the validation set.

\balance
\bibliographystyle{IEEEtran}
\bibliography{refs}

%
% or list them by yourself
% \begin{thebibliography}{9}
% 
% \bibitem{dcase2016web}
%   \url{http://www.cs.tut.fi/sgn/arg/dcase2016/}.
%
% \bibitem{IEEEPDFSpec}
%   {PDF} specification for {IEEE} {X}plore$^{\textregistered}$,
%   \url{http://www.ieee.org/portal/cms_docs/pubs/confstandards/pdfs/IEEE-PDF-SpecV401.pdf}.
%
% \bibitem{PDFOpenSourceTools}
%   Creating high resolution {PDF} files for book production with 
%   open source tools, 
%   \url{http://www.grassbook.org/neteler/highres_pdf.html}.
%
% \bibitem{eWilliams1999}
% E. Williams, \emph{Fourier Acoustics: Sound Radiation and Nearfield Acoustic
%   Holography}. London, UK: Academic Press, 1999.
% 
% \bibitem{ieeecopyright}
%   \url{http://www.ieee.org/web/publications/rights/copyrightmain.html}.
%
% \bibitem{cJones2003}
% C. Jones, A. Smith, and E. Roberts, ``A sample paper in conference
%   proceedings,'' in \emph{Proc. IEEE ICASSP}, vol. II, 2003, pp. 803--806.
% 
% \bibitem{aSmith2000}
% A. Smith, C. Jones, and E. Roberts, ``A sample paper in journals,'' 
%   \emph{IEEE Trans. Signal Process.}, vol. 62, pp. 291--294, Jan. 2000.
% 
% \end{thebibliography}

\end{sloppy}
\end{document}